\documentclass[a4paper]{jpconf}
\usepackage{graphicx}

\def\gtap{\;\raisebox{-.6ex}{\rlap{$\sim$}} \raisebox{.6ex}{$>$}\;}

\begin{document}
\title{
Quark-gluon plasma phenomenology from the lattice
}

\author{Chris Allton$^{1,\ast}$, Gert Aarts$^1$, Alessandro Amato$^{1,2}$,
  Wynne Evans$^{1,3}$, Pietro Giudice$^{1,4}$, Simon Hands$^1$, Aoife Kelly$^5$,
  Seyong Kim$^6$, Maria-Paola Lombardo$^7$, Sinead Ryan$^8$,
  Jon-Ivar Skullerud$^5$, Tim~Harris$^8$}

\address{
$^1$ Department of Physics, College of Science, Swansea University, Swansea SA2 8PP, U.K. \\
$^2$ Institute for Theoretical Physics, Universit\"at Regensburg,
  D-93040 Regensburg, Germany\\
$^3$ Fakult\"at f\"ur Physik, Universit\"at Bielefeld, D-33615
  Bielefeld, Germany \\
$^4$ Institut f\"ur Theoretische Physik, Universit\"at  M\"unster, Germany\\
$^5$ Department of Mathematical Physics, NUIM, 
     Maynooth, Co. Kildare, Ireland\\
$^6$ Department of Physics, Sejong University, Seoul 143-747, Korea\\
$^7$ INFN-Laboratori Nazionali di Frascati, I-00044, Frascati (RM) Italy\\
$^8$ School of Mathematics, Trinity College, Dublin 2, Ireland\\
$^\ast$ Speaker
}

\ead{c.allton@swan.ac.uk}

\begin{abstract}

The FASTSUM Collaboration has calculated several quantities relevant
for QCD studies at non-zero temperature using the lattice
technique. We report here our results for the (i) interquark potential
in charmonium; (ii) bottomonium spectral functions; and (iii)
electrical conductivity. All results were obtained with 2+1 flavours
of dynamical fermions on an anisotropic lattice which allows greater
resolution in the temporal direction.

\end{abstract}



\section{Introduction}
The Particle Data Book \cite{Beringer:1900zz} is a repository of
particle physics knowledge, and yet it contains no entries on the
deconfined phase of QCD. We present here some lattice calculations of
phenomena in the quark-gluon plasma (QGP) phase with the ultimate aim of
addressing this omission.

The lattice approach is based on the non-perturbative study of
correlation functions of operators calculated in a background of glue
and dynamical (sea) quarks using the imaginary time (i.e. Euclidean)
formulation\footnote{For a general introduction to lattice gauge
  theory see \cite{lgt-review}, and non-zero temperature reviews, see
  \cite{Levkova:2012jd,Lombardo:2012ix}.}.  For instance, to study the
$\eta_c$ system, correlators, $C(\tau)$, of the operator $J =
\overline{c} \gamma_5 c$ are determined ($c$ is the charm quark
field). Each state $i$ which has the same quantum numbers as $J$ (and
therefore can be excited by it) contributes a term $\sim \e^{-M_i
  \tau}$ to $C(\tau)$, where $M_i$ is the mass of state $i$. For large
$\tau$, only the ground state remains, and its properties can be
determined.

While the lattice technique has had many successes in calculating
phenomenologically relevant quantities at zero chemical potential,
$\mu$, particularly in the confined phase of QCD, it has a significant
limitation for $\mu>0$. This is the well-known ``sign problem'' which
has haunted efforts to extend our knowledge into the entire $(T,\mu)$
plane\footnote{For reviews of the sign problem, see \cite{sign}.}.

In this talk, I will discuss three lattice results obtained by
the FASTSUM Collaboration at $\mu=0$ and non-zero temperature, $T$, above and below
the deconfining temperature, $T_c$, all produced on
our 2+1 flavour, anisotropic lattices. These topics are a calculation
of the potential in the charmonium system (Sec.\ref{sec:charm_pot}),
bottomonium spectral functions (Sec.\ref{sec:bottom}), and a
determination of the electrical conductivity of QCD as a function of
$T$ (Sec.\ref{sec:conductivity}).

This work uses lattice simulations with the parameters in Table
\ref{tab:params} and a carefully crafted action with reduced lattice
artefacts \cite{Edwards:2008ja}. Anisotropic rather than isotropic
lattices provide a distinct advantage due to the finer temporal
resolution and correspondingly larger sampling of the correlator,
$C(\tau)$. This is particularly significant in the $T > 0$ case where
the temporal extent is limited to $0\le \tau \le 1/T$. Our $T_c$ value
is obtained from the Polyakov Loop.

\begin{table}[h]
\begin{center}
\begin{tabular}{crrrrrrrrrrrr}
\hline
$N_s$    &   32 &   24 &   24 &   24 &   32 &   24 &   32 &   24 &   32 &   24 &   24 &   32 \\
$N_\tau$ &   48 &   40 &   36 &   32 &   32 &   28 &   28 &   24 &   24 &   20 &   16 &   16 \\
$T$(MeV) &  117 &  141 &  156 &  176 &  176 &  201 &  201 &  235 &  235 &  281 &  352 &  352 \\
$T/T_c$  & 0.63 & 0.76 & 0.84 & 0.95 & 0.95 & 1.09 & 1.09 & 1.27 & 1.27 & 1.52 & 1.90 & 1.90 \\
\hline
\end{tabular}
\caption{\label{tab:params} Lattice parameters used where $N_{(s)\tau}$ is
  the number of sites in the spatial (temporal) direction.
Our spatial and temporal lattice spacings are $a_s = 0.1227(8)$ and
  $a_\tau = 0.0351(2)$ fm.  }
\end{center}
\end{table}



\section{Charmonium potential}
\label{sec:charm_pot}

The interquark potential in charmonium is of great interest to
phenomenologists modelling the QGP phase in heavy ion
collisions. Knowledge of this potential aids our understanding of
the $J/\psi$ system, and the extent to which
states become unbound with increasing temperature.

Lattice calculations of the finite temperature interquark potential
have, until recently, been restricted to the static (i.e. infinitely
heavy) quark limit \cite{static-pot}. On the other
hand, the finite-mass interquark potential has been calculated at
$T=0$ \cite{nbs} using the method developed by the
HAL QCD collaboration for internucleon potentials \cite{halqcd}.
This method first calculates the wavefunction of the two-particle system,
$\psi(r)$, which is then used as input into the Schr\"odinger
equation, yielding the potential, $V(r)$, as output.

In our case, we consider non-local operators of (charm)
quark fields, $J(x,\mathbf{r}) = c(x) \Gamma U(x,x+\mathbf{r}) 
\overline{c}(x+\mathbf{r})$, where $\Gamma$ is an
appropriately chosen Dirac (gamma) matrix, and $U(x,x+\mathbf{r})$ is
the gauge connection between $x$ and $x+\mathbf{r}$ \cite{ourWork}.
The correlation function of these operators is
\begin{equation}
C(\mathbf{r},\tau)
= \sum_\mathbf{x} \langle J(x,\mathbf{r}) J^\dag(x,\mathbf{0}) \rangle
= \sum_i 
\frac{\psi_i(\mathbf{r})\psi_i^\ast(\mathbf{0})}{2M_i} \e^{-M_i \tau},
\end{equation}
where the second sum is over the states $i$.
Following \cite{timedept},
$C(\mathbf{r},\tau)$ satisfies the Schr\"odinger equation which can then be used to
extract the potential.

\begin{figure}[h]
\begin{minipage}[t]{75mm}
\includegraphics[width=75mm]{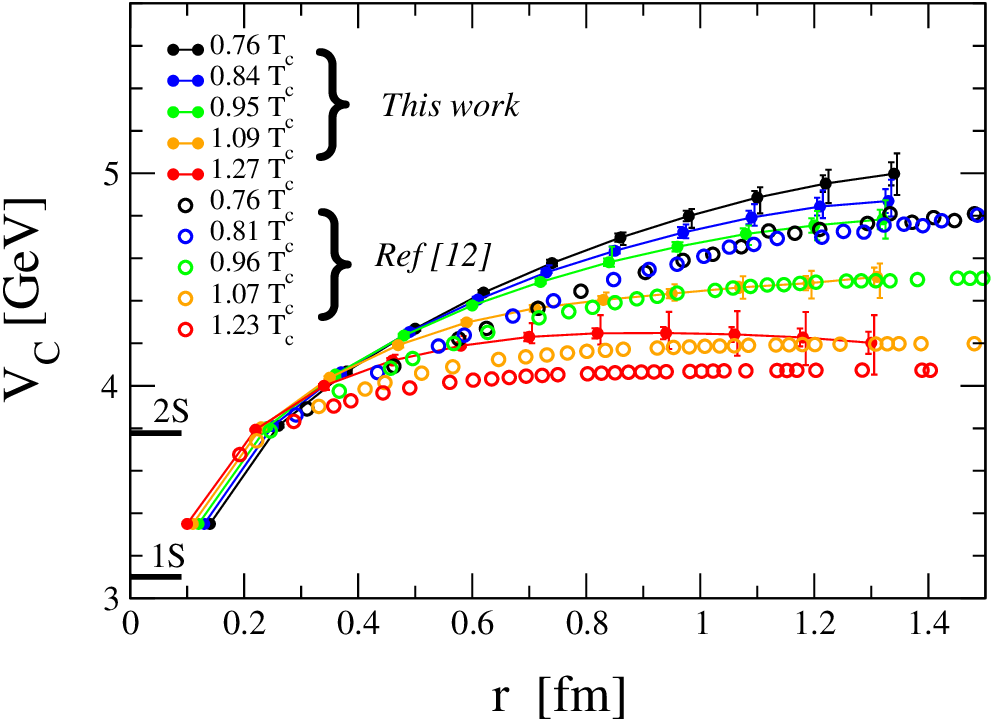}
\end{minipage}
\hspace{2pc}
\begin{minipage}[t]{75mm}
\includegraphics[width=75mm]{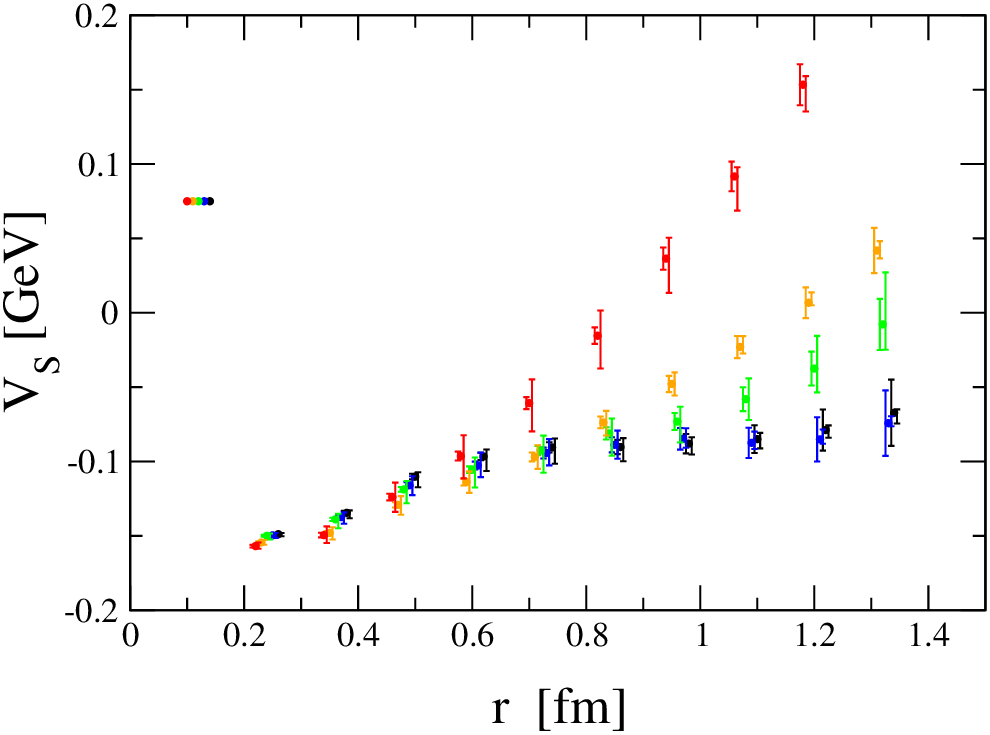}
\end{minipage}
\caption{\label{fig:V}
(Left) The charmonium central potential, $V_C(r)$,
  together with the results from a static quark calculation
  \cite{Kaczmarek:2005ui}. (Right) The charmonium spin dependent potential,
  $V_S(r)$.
}
\end{figure}

As usual, the potential can be decomposed into the central, $V_C(r)$,
and the spin dependent, $V_S(r)$, terms,
%
$V_\Gamma(r) = V_C(r) + \mathbf{s}_1\cdot \mathbf{s}_2 \;V_S(r)$,
%
where $\mathbf{s}_{1,2}$ are the spins of the charm quarks.  In
Fig.\ref{fig:V} (Left), we show the charmonium central potential,
$V_C(r)$ thus obtained at various temperatures above and below the
deconfining temperature. We use $N_s=24$ in this work. Also shown is
another determination of this potential, but obtained in the static
quark limit \cite{Kaczmarek:2005ui}. As can be seen, there is a clear
temperature dependence in our calculated potential which becomes less
confining as $T$ increases.  In Fig.\ref{fig:V} (Right) $V_S(r)$, is
plotted showing a repulsive core with a temperature variation at large
distances.

\section{Bottomonium spectral functions}
\label{sec:bottom}

In this section we discuss our studies of bottomonium at $T>0$.
Recent results from CMS show that the 1S and 2S/3S $\Upsilon$ states have
different relative multiplicities in Pb-Pb compared to p-p collisions
at the LHC \cite{Chatrchyan:2011pe}.  This result confirms the
picture, originally proposed in the charmonium system, in which the
higher mass states in quarkonium are the first to become unbound as
the temperature increases beyond $T_c$ \cite{Matsui:1986dk}.

We have performed lattice simulations of the bottomonium system using
an ${\cal O}(v^4)$ NRQCD lattice action \cite{Bodwin:1994jh} to
represent the $b$-quarks. This extends our earlier work \cite{ourNRQCD}.
NRQCD is an approximation obtained from QCD as a velocity expansion in
$v/c$, and is thus applicable for $b$-quarks.  The advantages of NRQCD
over the (exact) relativistic quark formulation are two-fold. There is
no periodicity in time which complicates the correlation function: in
the relativistic case, backward movers effectively half the number of
time points that carry independent information. The lack of these thermal
boundary effects means that the NRQCD quarks should be viewed as test
colour charges moving in a thermal bath of dynamical light quarks and
gluons.  Secondly, the solution of the NRQCD propagator is an initial
value problem which is much easier to solve than the matrix inversion
required in the relativistic case. Thus, for a given computer
resource, much higher statistics can be achieved in the NRQCD case.

The spectral function, $\rho(\omega)$, can be defined from the
two-point correlation function, $C(\tau)$, of bottomonium operators via
\begin{equation}
C(\tau) = \int \rho(\omega) K(\tau,\omega) d\omega,
\label{eq:rho}
\end{equation}
where in the NRQCD case the kernel is $K(\tau,\omega) = \exp(-\omega
\tau)$. In principle, the spectral function contains
complete information on the states in the channel considered. For a
stable particle of mass $M$, $\rho(\omega) \propto
\delta(\omega-M)$. For a resonance, this $\delta$-function
broadens acquiring a non-zero width, and if the state becomes
unbound, the spectral feature disappears.

\begin{figure}[h]
\begin{minipage}[t]{75mm}
\includegraphics[width=75mm]{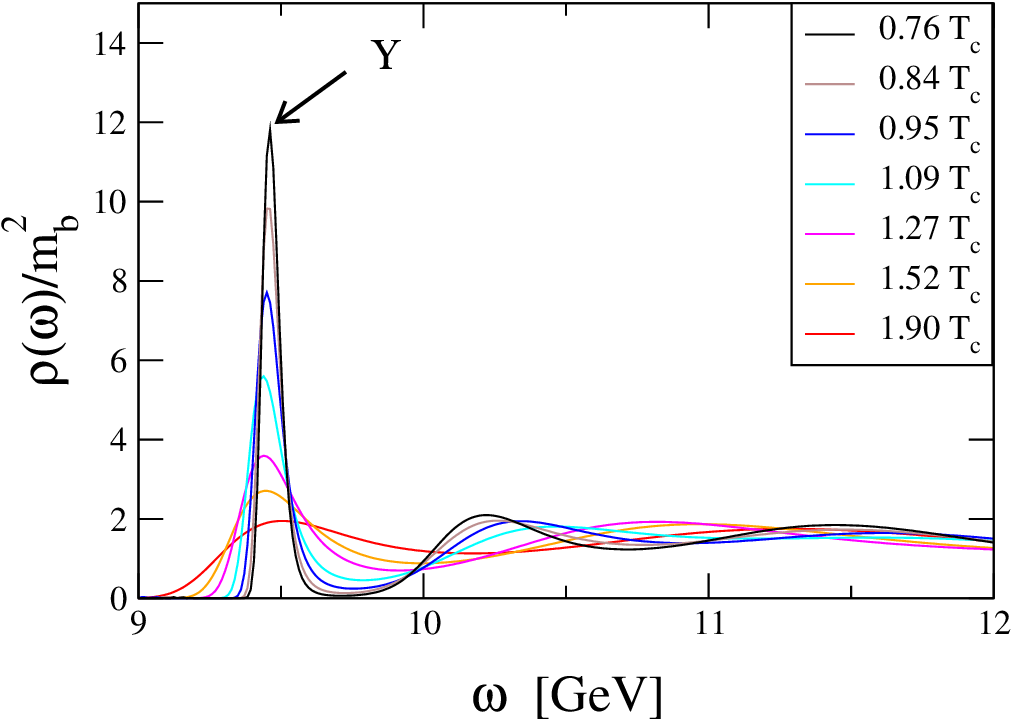}
\end{minipage}
\hspace{2pc}
\begin{minipage}[t]{75mm}
\includegraphics[width=75mm]{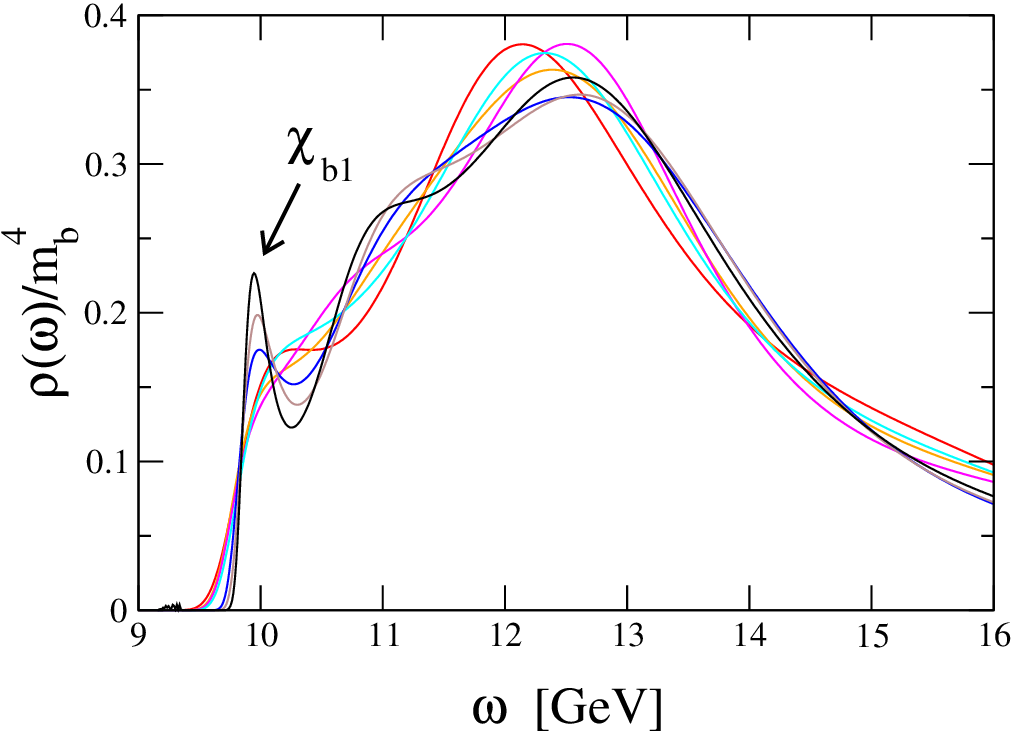}
\end{minipage}
\caption{\label{fig:bottom}
The NRQCD Bottomonium $S$-wave (Left) and $P$-wave (Right) spectral functions.}
\end{figure}

We have used the Maximum Entropy Method (MEM), a Bayesian technique,
to de-convolve eq.(\ref{eq:rho}) to extract $\rho(\omega)$
\cite{Asakawa:2000tr}.  The results for the $S$-wave ($\Upsilon$)
channel and $P$-wave ($\chi_{b1}$) channels are shown in
Fig.\ref{fig:bottom}, all obtained with $N_s=24$. There is a distinct
temperature dependence in both channels. While the $\Upsilon$ ground
state (1S) peak is seen to decrease with $T$, it remains a distinct
(i.e. bound) feature up to $T \approx 1.90 T_c$. However, the excited
state (2S) peak seems to disappear for $T\gtap T_c$. This agrees with
the experimental result obtained by the CMS collaboration
\cite{Chatrchyan:2011pe}. These results contrast with the $\chi_{b1}$
case where the ground state (1P) appears to melt at $T \approx T_c$,
see Fig.\ref{fig:bottom} (Right).

\section{Electrical conductivity}
\label{sec:conductivity}

\begin{figure}[h]
\begin{minipage}[t]{75mm}
\includegraphics[width=75mm]{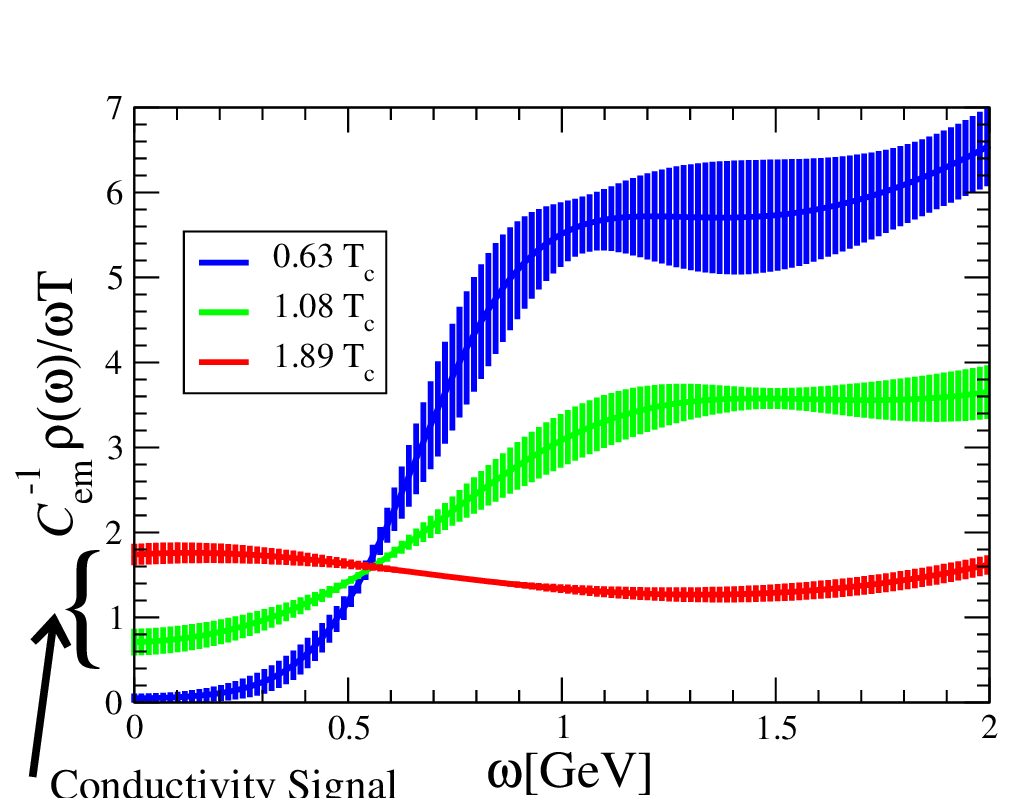}
\caption{\label{fig:transport}
The (light-quark) electromagnetic
spectral function, $\rho^{\rm em}(\omega)$, showing the conductivity
signal in the $\omega \rightarrow 0$ limit.
$C_{\rm em} = 5/9e^2$ for two light flavours.
}
\end{minipage}
\hspace{2pc}
\begin{minipage}[t]{75mm}
\includegraphics[width=75mm]{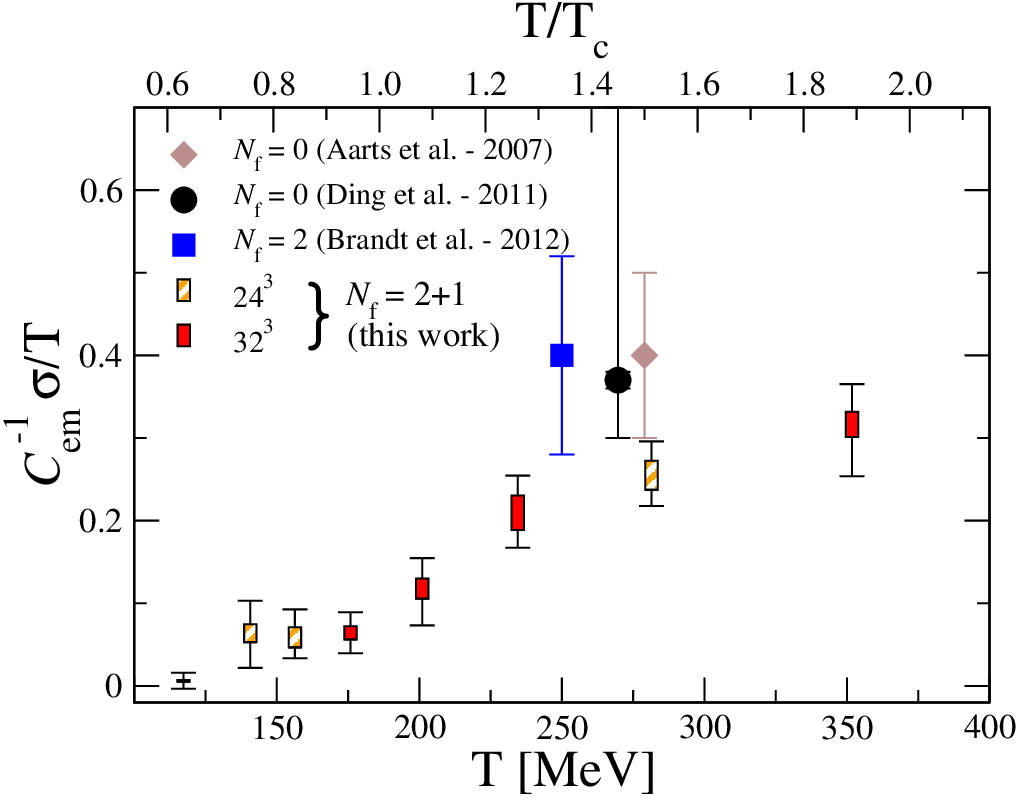}
\caption{\label{fig:conductivity}
The conductivity, $\sigma$, as a function of temperature. As well as
the results from our work, quenched ($N_f=0$)
\cite{Aarts:2007wj,Ding:2010ga} and dynamical results \cite{brandt}
are also shown.}
\end{minipage}
\end{figure}

We now discuss our calculation of the electrical conductivity,
$\sigma$ \cite{Amato:2013naa}. Transport coefficients such as $\sigma$
are of special relevance for understanding the bulk properties of the
fireball in relativistic heavy-ion collision experiments. Typically,
they can be determined from the zero energy limit of the appropriate
spectral function, i.e. $\lim_{\omega \rightarrow 0} \rho/\omega$.  In
the case of the conductivity, the correlator to consider is that of
the electromagnetic current of quark flavour $f$,
\begin{equation}
C^{\rm em}(\tau) = \sum_i \int d^3x
\langle j^{\rm em}_i(\tau,\mathbf{x})
        j^{\rm em}_i(0,\mathbf{x})^\dagger \rangle
\quad
\mbox{with }\quad
j^{\rm em}_i(x) = e\sum_f q_f j^f_i(x),
\end{equation}
where $e$ is the electron's charge, $q_f = \frac{2}{3},-\frac{1}{3}$,
is the fractional charge and $j^f_i \sim \overline{\psi}^f \gamma_i
\psi^f$ the number density current of quark field $\psi^f$.  In our
case \cite{Amato:2013naa} we use the conserved lattice vector current
(i.e. $\partial_\mu j^f_\mu = 0$) which means we do not have to
renormalise our results. We again use MEM to invert the correlator
to extract the corresponding $\rho^{\rm em}(\omega)$, see
Eq(\ref{eq:rho}), but this time with the relativistic kernel
$K(\tau,\omega) = \cosh[\omega(\tau-1/2T)] / \sinh[\omega/2T]$.

The conductivity is obtained via the Kubo relation,
\begin{equation}
\frac{\sigma}{T} = \frac{1}{6T}
\lim_{\omega \rightarrow 0} \frac{\rho^{\rm em}(\omega)}{\omega}.
\end{equation}

Fig.\ref{fig:transport} shows $\rho^{\rm em}(\omega)$
around $\omega \approx 0$ for three temperatures spanning $T_c$.
There is a clear non-zero
intercept at $\omega=0$ for $T\gtap T_c$, indicating a conductivity
signal. In Fig.\ref{fig:conductivity}, we present our $\sigma$ values
for all the $T$ values we studied. The dimensionless ratio $\sigma/T$
increases across the transition temperature.  Results from three other
published works are also shown \cite{Aarts:2007wj,Ding:2010ga,brandt}.



\section{Conclusion}
\label{sec:conclusion}

In this talk, I have summarised our collaboration's lattice calculations
showing that the charmonium potential becomes less binding with higher
$T$, that the $\Upsilon(1S)$ state survives above $T_c$ but the
$\Upsilon(2S)$ and $\chi_{b1}$ don't, and that the conductivity
increases with $T$ across the transition.





\section*{References}
\begin{thebibliography}{9}

\bibitem{Beringer:1900zz}
  J.~Beringer {\it et al.}  [Particle Data Group Collaboration],
  Phys.\ Rev.\ D {\bf 86} (2012) 010001.

\bibitem{lgt-review}
I.~Montvay, G.~M\"unster,
``Quantum Fields on a Lattice'', Cambridge Monographs on Mathematical
Physics.

\bibitem{Levkova:2012jd}
  L.~Levkova,
  PoS LATTICE {\bf 2011} (2011) 011
  [arXiv:1201.1516 [hep-lat]].

\bibitem{Lombardo:2012ix}
  M.~P.~Lombardo,
  PoS LATTICE {\bf 2012} (2012) 016
  [arXiv:1301.7324 [hep-lat]].

\bibitem{sign}
  P.~de Forcrand,
  PoS LAT {\bf 2009} (2009) 010
  [arXiv:1005.0539 [hep-lat]].
  G.~Aarts,
  PoS LATTICE {\bf 2012} (2012) 017
  [arXiv:1302.3028 [hep-lat]].


\bibitem{Edwards:2008ja}
  R.~G.~Edwards, B.~Joo and H.~-W.~Lin,
  Phys.\ Rev.\ D {\bf 78} (2008) 054501
  [arXiv:0803.3960 [hep-lat]].

\bibitem{static-pot}
  O.~Kaczmarek, F.~Karsch, F.~Zantow and P.~Petreczky,
  Phys.\ Rev.\ D {\bf 70}, 074505 (2004)
  [Erratum-ibid.\ D {\bf 72}, 059903 (2005)]
  [hep-lat/0406036],
  Y.~Maezawa {\it et al.}  [WHOT-QCD Collaboration],
  Phys.\ Rev.\ D {\bf 75}, 074501 (2007)
  [hep-lat/0702004],
  A.~M{\'o}csy and P.~Petreczky,
  Phys.\ Rev.\ D {\bf 77}, 014501 (2008)
  [arXiv:0705.2559 [hep-ph]],
  Z.~Fodor, A.~Jakov{\'a}c, S.~D.~Katz and K.~K.~Szabo,
  PoS LAT {\bf 2007}, 196 (2007)
  [arXiv:0710.4119 [hep-lat]],
  P.~Petreczky, C.~Miao and A.~M{\'o}csy,
  Nucl.\ Phys.\ A {\bf 855}, 125 (2011)
  [arXiv:1012.4433 [hep-ph]].
  A.~Bazavov and P.~Petreczky,
  [arXiv:1210.6314 [hep-lat]].
  Y.~Burnier and A.~Rothkopf,
  Phys.\ Rev.\ D {\bf 86}, 051503 (2012)
  [arXiv:1208.1899 [hep-ph]].

\bibitem{nbs}
  Y.~Ikeda and H.~Iida,
  PoS LATTICE {\bf 2010}, 143 (2010)
  [arXiv:1011.2866 [hep-lat]],
  T.~Kawanai and S.~Sasaki,
  Phys.\ Rev.\ Lett.\  {\bf 107}, 091601 (2011)
  [arXiv:1102.3246 [hep-lat]],
  Phys.\ Rev.\ D {\bf 85}, 091503 (2012)
  [arXiv:1110.0888 [hep-lat]],
  PoS LATTICE {\bf 2011}, 126 (2011)
  [arXiv:1111.0256 [hep-lat]],
  H.~Iida and Y.~Ikeda,
  PoS LATTICE {\bf 2011}, 195 (2011).

\bibitem{halqcd}
  N.~Ishii, S.~Aoki and T.~Hatsuda,
  Phys.\ Rev.\ Lett.\  {\bf 99}, 022001 (2007)
  [nucl-th/0611096],
  S.~Aoki, T.~Hatsuda and N.~Ishii,
  Prog.\ Theor.\ Phys.\  {\bf 123}, 89 (2010)
  [arXiv:0909.5585 [hep-lat]],
  S.~Aoki {\it et al.}  [HAL QCD Collaboration],
  [arXiv:1206.5088 [hep-lat]].
  N.~Ishii [HAL QCD Collaboration],
  PoS LATTICE {\bf 2011}, 160 (2011),
  N.~Ishii {\it et al.}  [HAL QCD Collaboration],
  Phys.\ Lett.\ B {\bf 712} (2012) 437
  [arXiv:1203.3642 [hep-lat]].

\bibitem{ourWork}
  C.~R.~Allton, P.~W.~M.~Evans and J.~-I.~Skullerud,
  PoS LATTICE {\bf 2012} (2012) 082
  [arXiv:1306.3140 [hep-lat]].
  P.~W.~M.~Evans, C.~R.~Allton and J.~-I.~Skullerud,
  arXiv:1303.5331 [hep-lat].
  P.~W.~M.~Evans, C.~Allton, P.~Giudice and J.~-I.~Skullerud,
  PoS LATTICE {\bf 2013} (2013) 168
  arXiv:1309.3415 [hep-lat].

\bibitem{timedept}
  N.~Ishii [HAL QCD Collaboration],
  PoS LATTICE {\bf 2011}, 160 (2011),
  N.~Ishii {\it et al.}  [HAL QCD Collaboration],
  Phys.\ Lett.\ B {\bf 712} (2012) 437
  [arXiv:1203.3642 [hep-lat]].

\bibitem{Kaczmarek:2005ui} 
  O.~Kaczmarek and F.~Zantow,
  Phys.\ Rev.\ D {\bf 71}, 114510 (2005)
  [hep-lat/0503017].

\bibitem{Chatrchyan:2011pe}
  S.~Chatrchyan {\it et al.}  [CMS Collaboration],
  Phys.\ Rev.\ Lett.\  {\bf 107} (2011) 052302
  [arXiv:1105.4894 [nucl-ex]].

\bibitem{Matsui:1986dk}
  T.~Matsui and H.~Satz,
  Phys.\ Lett.\ B {\bf 178}, 416 (1986).

\bibitem{Bodwin:1994jh}
  G.~T.~Bodwin, E.~Braaten and G.~P.~Lepage,
  Phys.\ Rev.\ D {\bf 51} (1995) 1125
   [Erratum-ibid.\ D {\bf 55} (1997) 5853]
  [hep-ph/9407339].

\bibitem{ourNRQCD}
  G.~Aarts, S.~Kim, M.~P.~Lombardo, M.~B.~Oktay, S.~M.~Ryan,
  D.~K.~Sinclair and J.~-I.~Skullerud,
  Phys.\ Rev.\ Lett.\  {\bf 106} (2011) 061602
  [arXiv:1010.3725 [hep-lat]],
  G.~Aarts, C.~Allton, S.~Kim, M.~P.~Lombardo, M.~B.~Oktay,
  S.~M.~Ryan, D.~K.~Sinclair and J.~I.~Skullerud,
  JHEP {\bf 1111} (2011) 103
  [arXiv:1109.4496 [hep-lat]],
  JHEP {\bf 1303} (2013) 084
  [arXiv:1210.2903 [hep-lat]].

\bibitem{Asakawa:2000tr}
  M.~Asakawa, T.~Hatsuda and Y.~Nakahara,
  Prog.\ Part.\ Nucl.\ Phys.\  {\bf 46} (2001) 459
  [hep-lat/0011040].

\bibitem{Amato:2013naa}
  A.~Amato, G.~Aarts, C.~Allton, P.~Giudice, S.~Hands and
  J.~-I.~Skullerud,
  arXiv:1307.6763 [hep-lat].

\bibitem{Aarts:2007wj}
  G.~Aarts, C.~Allton, J.~Foley, S.~Hands and S.~Kim,
  Phys.\ Rev.\ Lett.\  {\bf 99} (2007) 022002.

\bibitem{Ding:2010ga}
  H.~-T.~Ding {\it et al.},
  Phys.\ Rev.\ D {\bf 83} (2011) 034504.

\bibitem{brandt}
  B.~B.~Brandt, A.~Francis, H.~B.~Meyer and H.~Wittig,
  JHEP {\bf 1303} (2013) 100.

\end{thebibliography}
\end{document}